\documentclass[prb,twocolumn,showpacs,preprintnumbers,amsmath,amssymb]{revtex4-1}
\usepackage{float}
\usepackage{longtable}
\usepackage{dcolumn}
\usepackage{bm}
\usepackage{graphicx}
\usepackage{hyperref}
\usepackage[usenames]{color}
\usepackage{datetime}
\usepackage{color}
\usepackage{textcomp}
\usepackage{ulem}

\begin{document}

\title{Three-Dimensional Fermi-Surface and Electron-Phonon Coupling in Semimetallic 1$T$- TiTe$_2$ studied by Angle-Resolved Photoemission Spectroscopy}

\author{Xiao-Fang Tang$^{1,2}$}
\author{Yu-Xia Duan$^2$}
\author{Fan-Ying Wu$^2$}
\author{Shu-Yu Liu$^2$}
\author{Chen Zhang$^2$}
\author{Yin-Zou Zhao$^2$}
\author{Jiao-Jiao Song$^2$}
\author{Yang Luo$^2$}
\author{Qi-Yi Wu$^2$}
\author{Jun He$^2$}
\author{H. Y. Liu$^3$}
\author{Wen Xu$^{1,4}$} \email{Corresponding author: wenxu_issp@aliyun.com}
\author{Jian-Qiao Meng$^{2,5}$} \email{Corresponding author: jqmeng@csu.edu.cn}

\affiliation{$^1$Key Laboratory of Materials Physics, Institute of Solid State Physics,
Chinese Academy of Sciences, Hefei 230031, China}
\affiliation{$^2$School of Physics and Electronics, Central South University, Changsha 410083, Hunan, China}
\affiliation{$^3$Strong-field and Ultrafast Photonics Lab, Institute of Laser Engineering,
Beijing University of Technology, Beijing 100124, China}
\affiliation{$^4$School of Physics and Astronomy and Key Laboratory of Quantum Information of
Yunnan Province, Yunnan University, Kunming 650091, China}
\affiliation{$^5$Synergetic Innovation Center for Quantum Effects and Applications (SICQEA),
Hunan Normal University, Changsha 410081, China}

\date{\today}

\begin{abstract}

We present an investigation on electronic structure of 1$T$-TiTe$_2$ material via high-resolution angle-resolved photoemission spectroscopy (ARPES), utilizing tunable photon energy excitations. The typical semimetal-like electronic structure is observed and examined, where multiple hole pockets related to Te 5$p$ bands and one electron pockets related to Ti 3$d$ band are populated. The obtained results reveals i) a pronounced three-dimensional (3D) electronic structure of 1$T$-TiTe$_2$ with typical semi-metallic features, for both the Ti 3$d$ and the Te 5$p$ states; ii) multiple Fermi surface (FS) sheets and complex band structure; and iii) an obvious kink in dispersion at an energy of about 18 meV below the Fermi energy, the first experimental observation of a kink structure in 1$T$-TiTe$_2$, which may originate from electron-phonon coupling. These important and significant findings can help us to gain an in-depth understanding of the 3D electronic structure of semimetallic 1$T$- TiTe$_2$.

\end{abstract}

\maketitle
\maketitle

\section{Introduction}
The layered transition metal dichalcogenides (TMDCs) have a long history of research in physics and material science \cite{Greenaway1965, Murray1972, Lee1970, Naito1982}. In recent years, this time-honored system has received increasing attention, not only because of their exotic properties such as charge density waves \cite{Lee1970, Naito1982, JJYang2012, DWShen2008, TValla2004, PChen2017} and superconductivity \cite{TValla2004, DKang2015}, but also because of their extremely large and and non-saturating magnetotransport \cite{IPletikosic2014, MNAli2014} and topological properties \cite{Soluyanov2015, ZWang2016, XCPan2015, KZhang2016} and so on. In particular, TiTe$_2$ with structures related to the so-called 1$T$ polytype is a very interesting representative of the TMDCs. Up to date, TiTe$_2$ has been studied extensively owing to its interesting structural and electronic properties \cite{PChen2017, PBAllen1994, Koike1983, deBoer1984, Claessen1992, SHarm1994, JWAllen1995, Perfetti2001, Rossnagel2001, Perfetti2002, Nicolay2006, Strocov2006, FClerc2007, Krasovskii2007, Rossnagel2009, JKhan2012, VRajaji2018, UDutta2018}. TiTe$_2$ has been considered as a model material of Fermi liquid in the field of, e.g., high-resolution angle-resolved photoemission spectroscopy (ARPES) \cite{Claessen1992, SHarm1994, JWAllen1995, Perfetti2001, Rossnagel2001, Perfetti2002, Nicolay2006}. Most recent ARPES studies have shown that the single-layer TiTe$_2$ sample undergoes charge density wave (CDW) phase transition at 92$\pm$3K \cite{PChen2017}, which is not the case for bulk TiTe$_2$ \cite{deBoer1984, Koike1983, PBAllen1994}. More interestingly, nonhydrostatic pressure studies have found that TiTe$_2$ can even exhibit superconductivity \cite{UDutta2018}. Hence, the investigation of basic electronic structure and properties of 1$T$-TiTe$_2$ material is of great importance and significance in understanding quantum physics phenomena and in exploring potential applications in advanced electronic and optoelectronic devices.

As we know, most of the macroscopic properties of an electronic material are dictated by its microscopic band structure and electronic dynamics, specifically within an electronic energy regime of a few meV near the FS. It is therefore of great importance to reveals the topological change of the FS since it is intimately related to many of its low-energy properties such as electronic transport, specific heat, magnetic susceptibility, etc. Furthermore, the shape and size of the FS can be applied for determination of the carrier density, carrier type, carrier distribution, etc. From an experimental point of view, ARPES is a very effective tool to study the electronic structure of materials. This becomes the prime motivation for us to study the electronic structure of 1$T$-TiTe$_2$ material by using modern technique such as ARPES in the present study.

It should be noted that 1$T$-TiTe$_2$ has a sandwich-like material structure, where the electrons in different layers are attracted by the van der Waals forces. Due to the weak inter-layer bonding, high-quality TiTe$_2$ single crystals hold a promise of obtaining atomically flat and defect-free surface by cleaving under ultrahigh-vacuum. When 1$T$-TiTe$_2$ was considered as a textbook Fermi liquid material, its 3D electronic structures were often ignored. But in fact, even very anisotropic physical systems remain residual three-dimensionality. As has been pointed out that although the crystal structure of 1$T$-TiTe$_2$ is mainly two-dimensional (2D), its electronic structure can have very strong 3D-like natures \cite{Rossnagel2001, Strocov2006, Rossnagel2009}. The effects of dispersion along the $k_z$ are important and should be taken into account in analyzing and interpreting of the ARPES data. In this study, we intend to measure and examine the 3D electronic structure of 1$T$-TiTe$_2$ at relatively low temperatures. Through ARPES measurement via tunable synchrotron radiation with sufficient energy and momentum resolution, we would like to reveal the 3D electronic structure of 1$T$-TiTe$_2$ in different energy bands and to observe the corresponding FSs and complex band structure. In Section II, the samples and experimental measurement using ARPES are described briefly. The results obtained from this study are presented and discussed in Section III and the main conclusions are summarized in Section IV.

\section{samples and experimental measurement}

In this study, high-quality 1$T$-TiTe$_2$ single crystals were grown from the elements by chemical vapor transport using iodine as transport agent. The 1$T$-TiTe$_2$ samples used in the study were cleaved $in$ $situ$. The details of the sample preparation were documented elsewhere \cite{Rossnagel2009}.

For experimental measurement, high resolution ARPES measurements were carried out at beamline 5-4 of the Stanford Synchrotron Radiation Lightsource (SSRL), through using a Scienta R4000 electron energy analyzer. The measurements were carried out at a temperature of 10 K and in ultrahigh vacuum with base pressure better than 3$\times$10$^{-11}$ mabr. In the experiments, the wide angle $k_x$ - $k_z$ Fermi surface (FS) mapping were measured in the photon energy range from 16 eV to 30 eV in steps of 1 eV, with varying energy resolution from 8 meV to 21 meV. Phonton energy of 21 eV and 29 eV with corresponding energy resolution of 8 meV and 17 meV were used for probing of the $k_x$ - $k_y$ Fermi surface. Moreover, a light beam with a photon energy of 20 eV with an energy resolution of $\sim$ 10 meV was chosen to study the Ti 3$d$ band structure along high symmetry $\Gamma$(A)-M(L) direction.

\section{results and discussions}

\begin{figure}[bp]
\begin{center}
\includegraphics[width=0.95\columnwidth,angle=0]{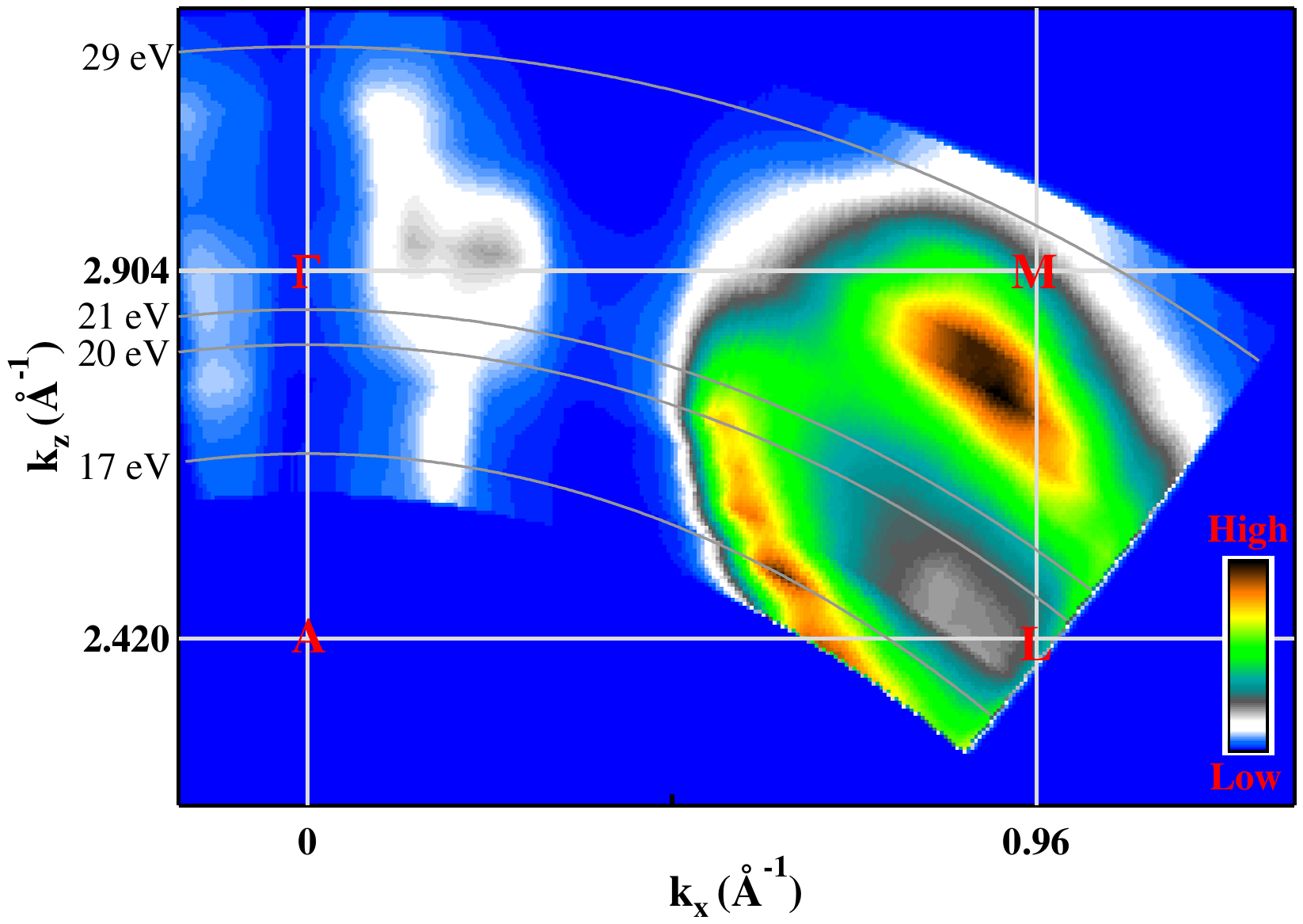}
\end{center}
\vspace*{-0.5cm} \caption{(color online) The map of three-dimensional Fermi surface of 1$T$-TiTe$_2$. The measurements are performed in the $\Gamma$$AML$ plane ($k_x$ - $k_z$). Different values of $k_z$ are accessed by varying the photon energy between 16 and 30 eV. The grey lines indicate the final state arc for different photon energies.}
\end{figure}

In the present study, we focus our attention on the 3D electronic structure of 1$T$-TiTe$_2$ at a relatively low temperature of 10 K. The 3D electronic band structure of 1$T$-TiTe$_2$ obtained from photon energy dependent normal emission is shown in Fig. 1. The measurements were performed in a section of the high-symmetry $\Gamma$$AML$ plane. For experimental convenience, the measurements were carried out with low photon energies, which holds high momentum and energy resolution. Different values of $k_z$ were accessed by varying the photon energy between 16 and 30 eV and estimated on the basis of an inner potential $V_0$ of 14.3 eV. As can be seen from Fig. 1, all the M(L) around Ti 3$d$ band and $\Gamma$(A) centered Te 5$p$ bands show very strong $k_z$ dispersion. The shape and intensity of the most important and widely studied Ti 3$d$ vary greatly with changing photon energy. The local-density approximations (LDA) calculation predict a 3D FS consisting of three hole pockets \cite{Strocov2006, Rossnagel2009, PChen2017}- two $\Gamma$-A centered corrugated cylinders and a $\Gamma$ centered lens-shaped FS \cite{Rossnagel2009}. Corrugated cylinder features can be found from the $k_x$-$k_z$ FS map. While, we cannot distinguish the small lens-shaped Te 5$p_z$ Fermi pocket, we can only see complex FS structures around $\Gamma$ point. That is probably because of the significant $k_z$ broadening \cite{VNStrocov2003, HWadati2006, Fujimori2016}. The strong 3D dispersion and the limited $k_z$ resolution result in photoemission features containing of signals from the nearby photons.

\begin{figure*}[tbp]
\begin{center}
\includegraphics[width=1.95\columnwidth,angle=0]{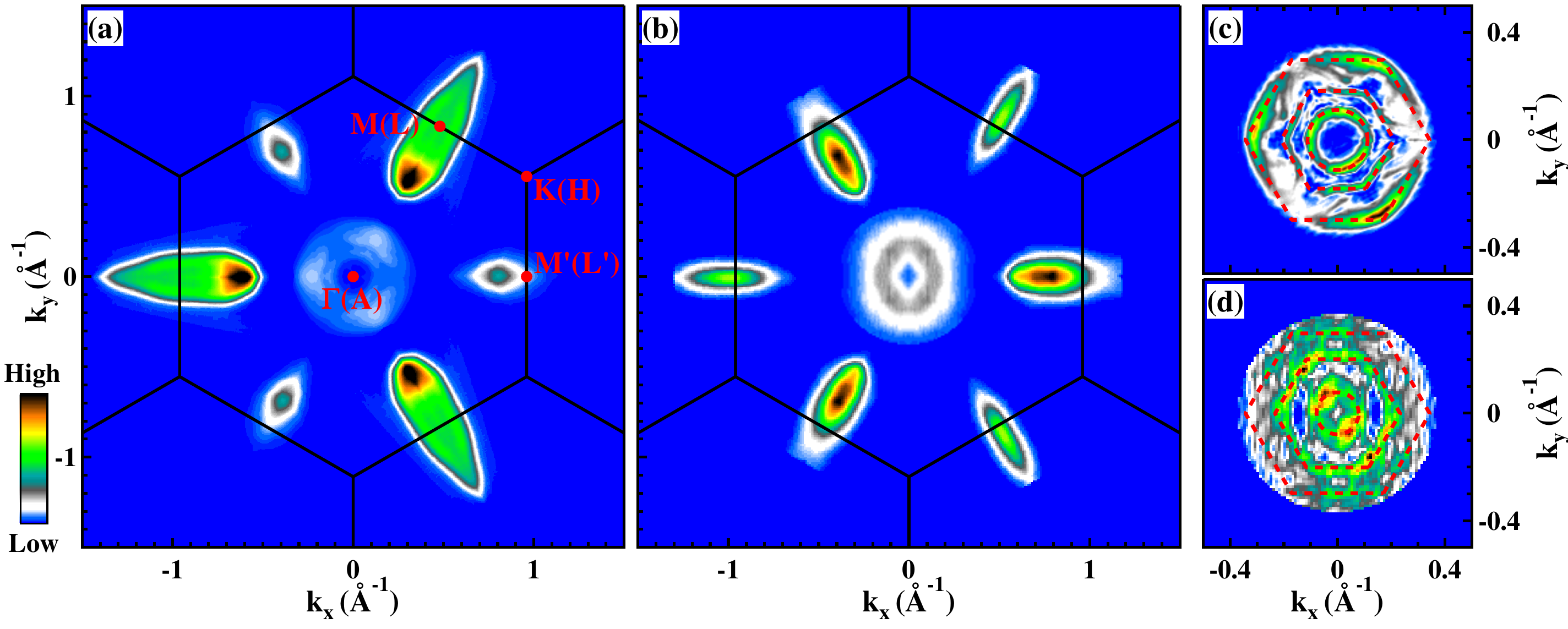}
\end{center}
\vspace*{-0.5cm} \caption{(color online) Two-dimensional Fermi surface mapping of 1$T$-TiTe$_2$ in the $k_x$-$k_y$ plane(symmetrized according to the crystal symmetry). The maps of photoemission intensity are taken with 21 eV [in (a)] and 29 eV [in (b)] photons, respectively. In (c) and (d), the FSs near $\Gamma$(A) point are treated with minimum gradient method corresponding to (a) and (b), respectively.}
\end{figure*}

In the present study, the constant photon energy FS mapping were undertaken to clarify the topological structure of the FS. Fig. 2(a) and (b) shows the FS mapping of the 1$T$-TiTe$_2$ with photon energy of 21 and 29 eV, respectively. The corresponding $k_z$ positions are indicated by grey lines in Fig. 1. These two energies are chosen for one energy to probe the FS near $\Gamma$ point and another energy to probe FS away from $\Gamma$ point and near M point. Superposed on the Fig. 2(a) and (b) are the Brilliouin zone with high-symmetry points indicated. All maps exhibit a trigonal symmetry duo to the space group $P$\={3}$m$1 of the 1$T$-type TMDCs. One can easily identify two types of the FS sheets: i) hole pockets centered at the ${\Gamma}$ point and derived from the Te 5$p$ bands and ii) the Ti 3$d$ band related to electron pockets centered at both the M(L) and M'(L') points. Noted that the strong 3D character of FS can be confirmed. The intensity, shape, and size of hole and electron pockets are changed significantly with photon energies: i) for 21 eV photon energy, the spectral weight of M(L)/M'(L') centered FS are high/weak, while for 29 eV it is just the opposite. ii) for 21 eV, the FS map show three high intensity drop-shaped features around M(L) points and three weaker elliptic features near M'(L') points. While for 29 eV, the FS map show elliptic features around all the M(L) and M'(L') points. iii) for 21eV, the size of M(L) centered electron pockets are much larger than that of M'(L') centered electron pockets. While for 29 eV photon energy, the size of M(L) centered electron pockets are significantly smaller than that of M'(L') centered electron pocket.

It can be found from Fig. 2 that, in consistent with previous calculations \cite{Strocov2006, Rossnagel2009, PChen2017}, the structure of $\Gamma$ centered hole pockets is very complex. Here we employ the minimum gradient method to examine the multiple hole pockets around $\Gamma$(A) point. This approach is supposed to be able to sharpen the weak structure \cite{YuHe2017}. The results are shown in Fig. 2 (c) and (d), corresponding to Fig. 2(a) and (b) respectively. The dashed red lines indicate the possible position of the FS. For the cases of both 21 eV and 29 eV photon energies, at least three distinct $\Gamma$(A) centered hole pockets can be resolved. The inner hole pocket is a circle and the outer two hole pockets are hexagon. On closer inspection, some very weak and shapeless features can be found between adjacent hole pockets (see Fig. 2(c)). The following Te 5$p$ band structures also confirm this. These findings are probably due to finite $k_z$ dispersion and limited $k_z$ resolution. However the most recent studies report another possibility, which suggests the ARPES contour maps should reveal band continuums \cite{PChen2017}.

Because the FS topologies around the $\Gamma$(A) point are complex and have strong 3D characteristics, it is very important to study the band structure at and away from the $\Gamma$ point. Fig. 3(a) and (b) show the ARPES images of photoelectron intensity as a function of energy and momentum for TiTe$_2$. Fig. 3(c) and (d) are 2D second-derivative images from the original data. We note that a typical semi-metal structure can be found in TiTe$_2$, namely a single Fermi level goes across the electron and valence-electron (or hole) bands so that the electrons and holes coexist in the material. The light Te 5$p$ bands approach the $\Gamma$(A) point and form the hole pockets, whereas the shallow Ti 3$d$ band forms an electron pocket near the M(L) point. Fig. 3(a) and (c), corresponding to 17 eV photon energy, away from $\Gamma$ point, clearly show two separated light Te 5$p$ bands approach the $\Gamma$(A) point and form the hole pockets and a shallow Ti 3$d$ band forms an electron pocket near the L point. For 21 eV photon excitation, close to the $\Gamma$ point, the band structures are more complex. In Fig. 3(b), two broad Te 5\emph{p} related humps are observed. It is obvious from the illustrated momentum distribution curve (MDC) at Fermi energy that it cannot be fitted with two peaks Lorentz lineshape. In Fig. 3(d), at high as 5 Te 5\emph{p} bands and a Ti 3$d$ band can be identified from the 2D second-derivative image. These features are different from the previous band structure calculations, which suggest three Te 5$p$ bands cross the Fermi level along the $\Gamma$-M direction \cite{Strocov2006, Rossnagel2009}. Our results are similar to those predicted by recent articles which pointed out the band should be continuums around the $\Gamma$ point \cite{PChen2017}.

\begin{figure}[tbp]
\begin{center}
\includegraphics[width=0.95\columnwidth,angle=0]{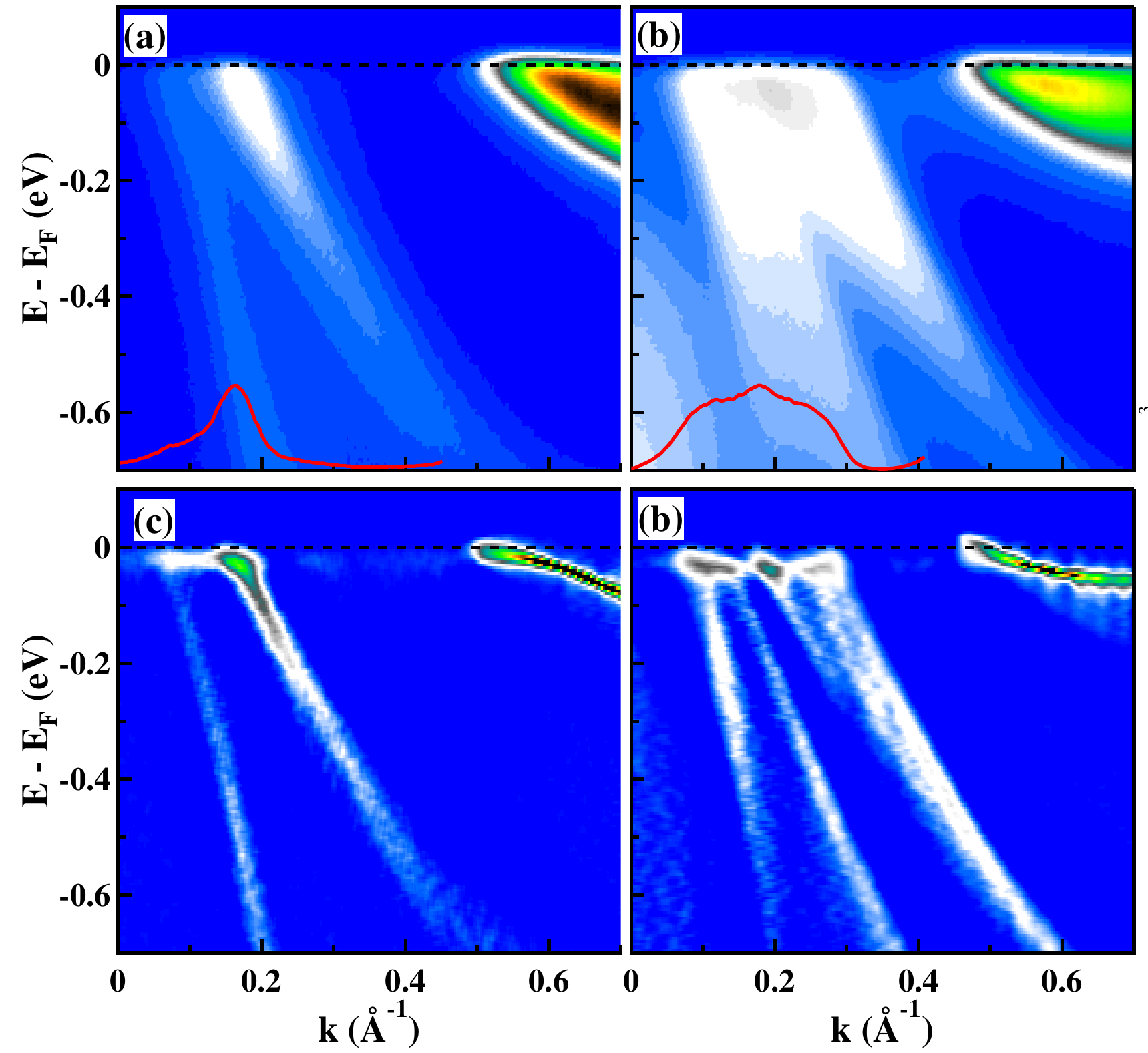}
\end{center}
\vspace*{-0.5cm} \caption{(color online) ARPES maps around $\Gamma$(A) at the zone center of 1$T$-TiTe$_2$. (a) and (b) ARPES maps, taken along the $\Gamma$(A)-M(L) direction with 17 and 21 eV photon energy. The energy resolution was set to 8 and 10 meV, respectively. The inset red lines represent the MDCs at Fermi energy of Ti 5$p$ bands. The corresponding $k_z$ position are indicated by grey lines in Fig. 1. (c) and (d) measured band structures reduced with the 2D second-derivative to enhance the weak bands while maintaining band dispersion.}
\end{figure}

Now we examine the features of the Ti 3$d$ band. Fig. 4(a) shows the high-resolution ARPES spectra of Ti 3$d$ band along the $\Gamma$(A)-M(L) direction with 20 eV photon energy. This photon energy was chosen to obtain very narrow quasiparticle bandwidth, where its final state arc intersects Fermi pocket in a nearly vertical way (see Fig. 1). As been discussed previously \cite{Claessen1992, SHarm1994, JWAllen1995, Perfetti2001, Rossnagel2001, Perfetti2002, Nicolay2006}, the momentum distribution curves (MDCs) of shallowed Ti 3$d$ band can no longer be fitted with Lorentzian lineshapes. An empirical approach is used to obtain its qualitatively dispersion. Fig. 4(b) shows the energy distribution curves(EDCs) band, which was extracted by taking second-derivative of original ARPES data (Fig. 4(a)) with respect to energy. One can see an obvious kink in dispersion $\sim$ 18 meV below Fermi energy, indicated by a red arrow. The original data (Fig. 4(a)) is divided by the Fermi distribution function to visually inspect the kink. A famous double peak (peak-dip-hump) feature can be observed in the Fig. 4(d), which were widely observed in other materials \cite{Hengsberger1999, LaShell2000, Lanzara2001, Kaminski2001}. This is the first experimental observation of a kink structure in 1$T$-TiTe$_2$. However, the kink structure is widespread in other materials, such as Bi (1000) surface \cite{Hengsberger1999, LaShell2000}, High-T$_c$ superconductor \cite{Lanzara2001, Kaminski2001, Bogdanov2000, PDJohnson2001}. The kink structure was also observed in other TMDs with CDW phase \cite{TValla2000, TValla2004}, but not in materials without CDW phase \cite{DWShenPhdThesis}. These features can be understood within a many-body framework via considering the electron interactions with impurities, phonons and other electrons. After comparing the experimental results obtained from the present photoemission measurements with those from Raman spectroscopy \cite{Hangyo1983, JKhan2012, VRajaji2018} and specific heat \cite{Rossnagel2009}, we believe that the kink in the energy dispersion in TiTe$_2$ is induced by electron-phonon coupling. This mechanism is proposed on the basis of following experimental findings. i) A mode ($A$$_1g$) regarding to electron-phonon coupling near 18 meV has been found in the previous Raman spectroscopy studies \cite{Hangyo1983, JKhan2012, VRajaji2018}; ii) The most recent result obtained from specific heat measurement indicates that the phonon effect corresponds to a Debye energy of about 19 meV \cite{Rossnagel2009}; And iii) our ARPES results shown in Fig. 4 indicate that the kink is observed in the low-energy regime around the Fermi-level, which is around the optic-phonon energy measured from Fermi energy in TiTe$_2$.

\begin{figure}[tbp]
\begin{center}
\includegraphics[width=0.95\columnwidth,angle=0]{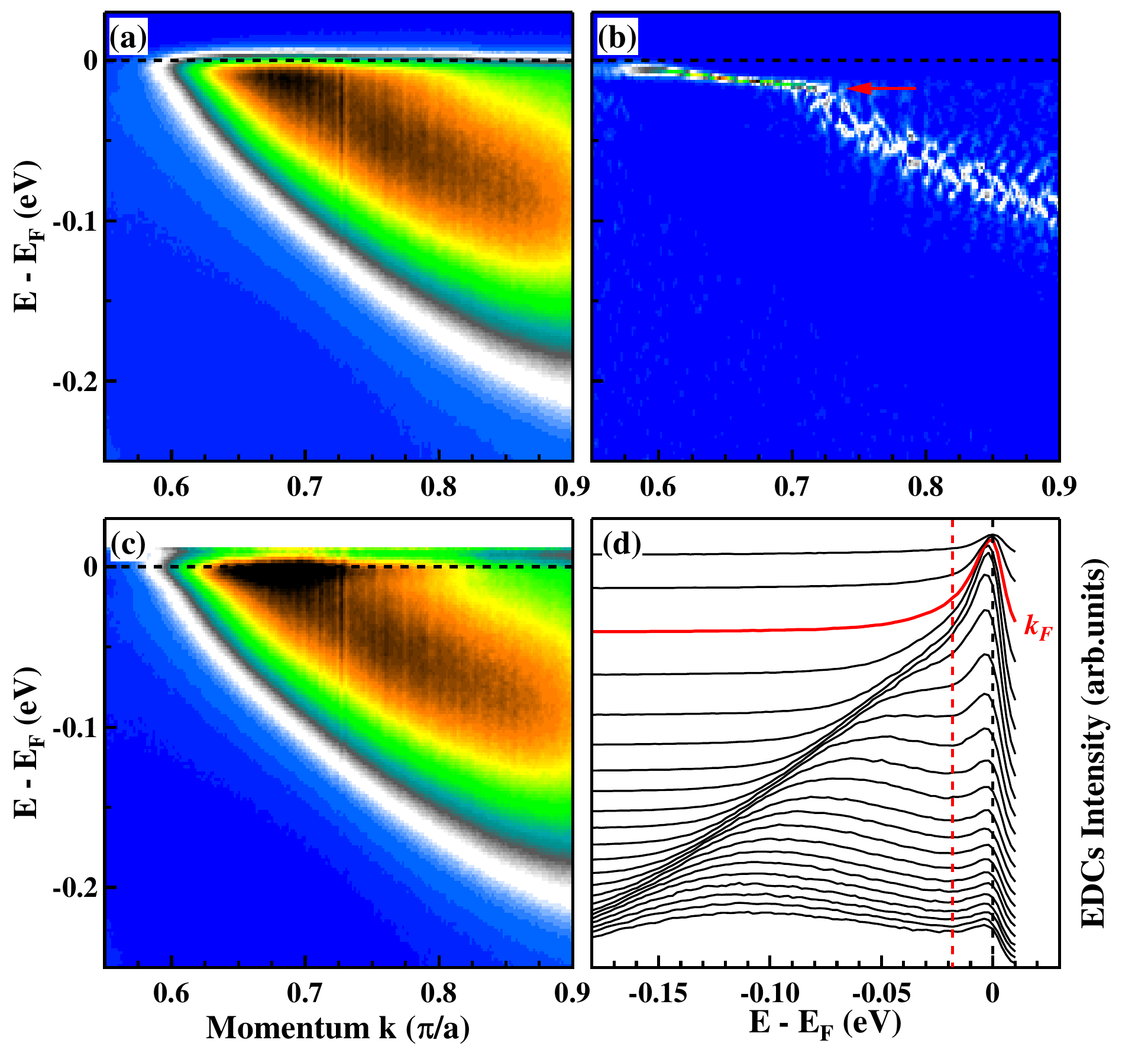}
\end{center}
\vspace*{-0.5cm} \caption{(color online) ARPES maps for Ti 3$d$ band of 1$T$-TiTe$_2$.(a) band structure plot along $\Gamma$(A)-M(L) direction with 20 eV photon energy. The angle and energy resolution was set to 0.2$^{\circ}$ and $\sim$7 meV, respectively. (b) measured structure reduced with the second-derivative method to enhance the weak bands. The red arrow marks the kink position.(c) measured image divided by the Fermi distribution function to highlight the kink feature.(d) EDCs corresponding to (c). The red dashed line indicated the position of dip.}
\end{figure}

TiTe$_2$ has often been considered as a testing material of Fermi liquid electron gas. However, its 3D features of the electronic band structures are often ignored when spectra are fitted with a Fermi liquid lineshape. The experimental results shown and discussed in this work indicate that although the crystal structure of 1$T$-TiTe$_2$ is mainly 2D-like, its electronic band structure can display very strong 3D-like natures, namely the electronic energy spectrum of 1$T$-TiTe$_2$ depends not only on the in-plane electron wavevector $(k_x,k_y)$ but also on $k_z$ the electron wavevector along the normal direction. A strong dependence of the electronic energy spectrum upon $k_z$ in 1$T$-TiTe$_2$ is basically a consequence that it is with a sandwich-like material structure in which the electrons in different layers are attracted by the van der Waals forces. Due to inter-layer bonding and corresponding electronic interactions, the electronic band structure along the $k_z$ or normal direction should be a functional form of the electronic wavevector or momentum along this direction.

\section{Conclusions}

In this work, we have investigated the low-energy electronic structure of the layered TMDCs 1$T$-TiTe$_2$ single crystal at a relatively low temperature using state-of-the-art technique such as the ARPES. The major aim of this study is at measuring and examining the 3D features of the electronic band structure of 1$T$-TiTe$_2$ material. The main conclusions deduced from the present study are as follows. i) The electronic band structure regarding to both the Ti 3$d$ and the Te 5$p$ states show a strong 3D topology. This result demonstrates that although the crystal structure of 1$T$-TiTe$_2$ is mainly 2D-like, its electronic band structure is very much 3D-like; ii) A typical semi-metal structure can be found in TiTe$_2$ in which a single Fermi level goes across the electron and hole bands so that the electrons and holes can coexist in the material; iii) The multiple $\Gamma$(A) centered hole pockets formed by Te 5$p$ bands and three M(M') around shallow electron pockets formed by Ti 3$d$ band have been experimentally identified. This finding allows us to give a more detailed description of the electronic structure of 1$T$-TiTe$_2$ single crystal; And iv) a kink can be observed in low photon energy regime at approximately 18 meV in Ti 3$d$ band. This feature is induced mainly by electron-phonon coupling in the material. We hope these important and significant experimental findings can help us to gain an in-depth understanding of the basic electronic structure of 1$T$-TiTe$_2$ which has been considered as a testing material for Fermi liquid electron gas.

\section{Acknowledgements}
We thank Gey-Hong Gweon for his unreserved support and Kai Rossnagel for high quality samples and constructive criticism. This work was supported by the National Natural Science Foundation of China (NSFC, Grants No.11574402 and No.51502351) and by the Innovation-driven Plan in Central South University (2016CXS032). WX was supported by the NSFC (Grant No. 11574319) and the Center of Science and Technology of Hefei Academy of Science (Grant No. 2016FXZY002).

\end{document}